\documentclass[11pt]{article}
\usepackage{amsmath}
\usepackage{amssymb}
\usepackage{fancyvrb}
\usepackage{footnote}
\usepackage{graphicx}
\usepackage{hyperref}

\ifx\keywords\undefined
    \providecommand{\keywords}[1]
    {
      \small	
      \textit{Keywords --} #1
    }

\newcommand{\Cpp}{C$^{\textnormal{\tiny ++}}$}
\newcommand{\JSON}{J{\scriptsize SON}}
\newcommand{\OBJS}{OB{\scriptsize JS}}
\newcommand{\POSIX}{P{\scriptsize OSIX}}
\newcommand{\PHP}{P{\scriptsize HP}}

\newtheorem{remark}{Remark}

\begin{document}

\title{Advanced OOP and new syntax patterns for Javascript}
\author{Alessandro Rosa\footnote{Freelance developer}}
\maketitle


\begin{abstract}
We present \OBJS, a new transpiler project featuring the implementation of
typified variables and functions call management in Javascript, as well as
several new operators and syntax patterns that could make coding more agile
and versatile. The goal is to empower this language. According to this point
of view, this transpiler aims at implementing Object Oriented Programming
paradigms into Javascript. The author opines that this would be likely the
best evolution of this language in ways that should be proper to the original
syntax, that is, by adopting native \Cpp standards, so that there would be no
promiscuity between old and new patterns, benefiting those who come from
similar languages.
\end{abstract}

\tableofcontents

\begin{flushright}
	`\emph{Words are not only tools to convey the thought,\\they are mainly
conditions for thinking}'\\
    \vspace{0.1cm}
	Martin Heidegger (1889–-1976)
\end{flushright}

\section{Introduction}
Javascript first appeared in the mid 1990s and has since become one of the most
widely used programming languages for front--end services of Web applications,
at both small and large scale projects. The increasing popularity of Javascript
has propelled it to the forefront of front-end services at any scale. In line
with the growing ambitions shown by recent systemic applications, the original
Javascript requires some enhancements in order to meet a broader range of
demands. The development of Typescript demonstrated that there are strong
motivations and consistent margins for the implementation of new patterns that
are further presented and pertain to areas that the original Javascript has not
yet covered. We presented some new ideas for the promotion of new coding
paradigms, inherent to advanced Object Oriented Programming (OOP) or the
`\emph{code event response}', based on our vision of code as a timeline.
Furthermore, the ability of `\emph{speaking in Javascript}' to develop custom
dialects in order to lower the entry level to code writing, especially for
beginners. These new features have been implemented into the transpiler
\OBJS.\footnote{The acronym of `\texttt{OB}ject \texttt{J}ava\texttt{S}cript'.}

\section{Disclaimer}\label{Disclaimer}
This paper is not intended to be an handbook of commands. The primary goal is
to illustrate the underlying paradigms. The audience is not expected to be
familiar with Javascript, but rather to understand the fundamentals of coding.

\section{A note about this project}\label{a_note_about}
This transpiler project has been developed in Javascript and it is currently
running as a demo
version\footnote{\url{http://alessandrorosa.altervista.org/objs/}} at the time
we are writing, as only formal parsing is performed: the (desired) full
functionalities cannot be granted, because Javascript does not feature a
sophisticated and granular memory management. For example, object typification
could not be fully resolved if depending on the evaluation of decisions at
runtime:

\begin{Verbatim}[fontsize=\scriptsize,numbers=none,frame=single]
complex a = (1,2), b = (2,1);
var c = 1, d = 2;

var _rnd = Math.random();
var _ret = _rnd < 0.5 ? a + b : c + d;
var _prod = _ret * 2;
\end{Verbatim}

While the formal approach allows to determine the correct behavior of the
operator `\texttt{+}' in terms of arguments datatype (row 5), it could not be
decided whether \texttt{\_ret} (row 5) represents the sum of two
\texttt{Number}s or of \texttt{complex} objects: in fact, the choice depends
upon the joint combination of the runtime value of \texttt{\_rnd} (row 4) with
inequality \texttt{\_rnd $<$ 0.5} (row 5), ultimately affecting the overload of
the operator `\texttt{*}' (row 6).

This transpiler project would like to be implemented in a single language, or
the individual features should be incorporated into a Javascript engine.

\section{New and revisited syntax patterns}
\subsection{Multiple actions}\label{subsection_multipleactions}
The abstract pattern of `multiple actions' is \texttt{<objects><op><values>}
and it extends the action of the abstract binary operator \texttt{<op>}, by
picking left and right values and dissecting the original pattern into a series
of the simpler \texttt{<object><op><value>}, by means of two indexed sequences
of objects. According to the number of right values, the relations, involved by
this pattern, could show up as \emph{one--to--many} or \emph{many--to--many}.

\subsubsection{Assignments.} Let the token \texttt{<op>} be the standard assignment operator
`\texttt{=}', in order to assign values to multiple variables. So the usage of
this binary operator extends, thanks to this pattern, the standard
destructuration. The multiple action here works in a \emph{one--to--many}
fashion:

\begin{Verbatim}[fontsize=\scriptsize,numbers=none,frame=single]
var obj0 = "str";
var (obj1, obj2, obj3) = obj0;
console.log( obj0, ">>", obj1, obj2, obj3 );
\end{Verbatim}

\noindent Or let this assignment in \emph{many--to--many} fashion:

\begin{Verbatim}[fontsize=\scriptsize,numbers=none,frame=single]
var (obj1, obj2) = (1, 2);
console.log( obj1, obj2 );
\end{Verbatim}
Here a simpler but nifty example. Suppose we need an array with subsets of
elements with a same value, according to a given law, here, the identity of the
index \texttt{\_i}, for sake of simplicity:

\begin{Verbatim}[fontsize=\scriptsize,numbers=none,frame=single]
var _arr = [], _limit = 2;
    for( var _i = 0 ; _i < _limit ; _i++ )
    {
        ( _arr[_i*_limit], _arr[_i*_limit+1], _arr[_i*_limit+2] ) = _i ;
    }

console.log( _arr );
\end{Verbatim}
We will get an array with nine entries, distributed as follows:
\texttt{[0,0,0,1,1,1,2,2,2]}. This operator performs multiple assignments and
works on the index that lists the appearance of each element on the left. The
index management allows to work with the concept of wildcard: the code below
includes the \emph{asterisk} token on the right, which encodes a special
assignment behavior: when the number of values on the left is greater than the
ones on the right.

\begin{Verbatim}[fontsize=\scriptsize,numbers=none,frame=single]
var (obj1, obj2, obj3, obj4) = (1, 2, *);
console.log( obj1, obj2, obj3, obj4 );
\end{Verbatim}
The value appearing immediately before the asterisk is assumed for filling the
rest of the left objects whose index places do not match the ones of the
elements on the right. In the above example, values will be assigned as
follows: 1 and 2 to \texttt{obj1} and \texttt{obj2} respectively; 2 to
\texttt{obj3} and to \texttt{obj4}. Not only basic datatypes can be assigned;
for example, the return value of the function \texttt{doSomething} too:

\begin{Verbatim}[fontsize=\scriptsize,numbers=none,frame=single]
function doSomething(a) {return a;}

var (obj1, obj2) = (1, doSomething(1));
console.log(obj2);
\end{Verbatim}
\noindent Or the contents of the function object itself:

\begin{Verbatim}[fontsize=\scriptsize,numbers=none,frame=single]
function doSomething(a) {return a;}

var (obj1, obj2) = (1, doSomething);
console.log( obj2 );
\end{Verbatim}
The assignment operation requires that the left containers would be eligible to
be filled with new values. This may not always be the case in the following
examples including \emph{not eligible containers}:

\begin{Verbatim}[fontsize=\scriptsize,numbers=none,frame=single]
(obj1, 1+2) = (1,2) // no formulas allowed on the left
(obj1, 1) = (1,2) // no explicit constant values
(obj1, var) = (1,2) // no reserved words
(obj1, 'string') = (1,2) // no strings
(obj1, _obj2.method) = (1,2) // no method objects
(obj1, doNothing(1)) = (1,2); // no function calls

const _a = 1;
(obj1, _a) = (1,2) // no constant variables

function doNothing() {}
(obj1, doNothing) = (1,2); //no functions allowed
\end{Verbatim}

\subsubsection{Comparisons.} This new pattern, in the basic assignment usage, may resemble
a slight evolution of the standard destructuring syntax. We will show the new
opportunities that arise when a new operator comes into play. We changed the
abstract token \texttt{<op>} to the loose comparison operator `\texttt{==}'.
This code in the \texttt{if} condition

\begin{Verbatim}[fontsize=\scriptsize,numbers=none,frame=single]
if ( obj1 == 0 && obj2 == 1 && obj3 == 2 )
{
    do_something();
}
\end{Verbatim}
\noindent can be crammed into this more compact version:

\begin{Verbatim}[fontsize=\scriptsize,numbers=none,frame=single]
if ( ( obj1, obj2, obj3 ) == ( 0, 1, 2 ) )
{
    do_something();
}
\end{Verbatim}
The new presentation appears to be a clearer version of the same instructions:
all objects that play the same role are grouped and separated on the same side,
reducing symbolic redundancy and the difficulty of reading a chain of different
symbols with different precedence, as shown above with `\texttt{\&\&}' (logic
and) and `\texttt{==}' (non--strict comparison).

\subsection{Array}
\noindent We will show below the patterns for special operations with arrays:

\subsubsection{Special extraction.} The following in line aim at extracting elements
from the original array according to the index range, which could consist of
singletons (each referring to one element only) or with different
(well--ordered) ends.

\begin{Verbatim}[fontsize=\scriptsize,numbers=none,frame=single]
var _a = [9,8,7,6,5,4,3,2,1,0];
var _a1 = _a[6:5:0:1]; // ':' only the input indexes
var _a2 = _a[3-->]; // '-->' from index 3 (included) to the end
var _a3 = _a[<--5]; // '<--' from index 0 to 5 (included)
var _a4 = _a[3>--<5]; // '>--<' from index 3 to 5 (both ends)
var _a5 = _a[2<-->7]; // '<-->' from begin to index 3 and from 5 to end (indexes
included)

console.log(_a1);
console.log(_a2);
console.log(_a3);
console.log(_a4);
console.log(_a5);
\end{Verbatim}

\subsubsection{Pushing an element once or repeatedly.}
When the token `\texttt{[]}' (opening and closing square brackets) is appended
to the identifier of an \texttt{Array} datatype then it turns into an operator
that pushes (i.e., it appends to the end of the array) one or more elements
inside it. We pushed only one element here below:

\begin{Verbatim}[fontsize=\scriptsize,numbers=none,frame=single]
var _a = [0];
_a[] = 2 ;
console.log( _a );
\end{Verbatim}
This pushing operator `\texttt{[]}' can be run multiple times in succession
when appearing in combination with the asterisk operator `\texttt{*}', counting
the value on the right, the container `\texttt{\_i}' here :

\begin{Verbatim}[fontsize=\scriptsize,numbers=none,frame=single]
var _a = [0], _i = 10;
_a[] * _i = 2 ;
console.log( _a );
\end{Verbatim}
\noindent The array `\texttt{\_a}' is filled in below with integer values
ranging from 0 to 9:

\begin{Verbatim}[fontsize=\scriptsize,numbers=none,frame=single]
var _a = [];
for( var _i = 0 ; _i < 10; _i++ ) _a[] = _i ;
console.log( _a );
\end{Verbatim}

\subsubsection{Popping one element once or repeatedly.}
The concepts behind the previous subsection can be borrowed from the popping
(i.e., removing from the end of the array) operator `\texttt{][}' (closing and
opening square brackets). In this case, however the pattern is simpler, because
there is no input value to append.

\begin{Verbatim}[fontsize=\scriptsize,numbers=none,frame=single]
var _a = [0,1,2,3];
_a][ ;
console.log( _a );
\end{Verbatim}
\noindent The following code pops out all elements from the array \texttt{\_a}:

\begin{Verbatim}[fontsize=\scriptsize,numbers=none,frame=single]
var _a = [0,1,2,3];
_a][ * 4 ;
console.log( _a );
\end{Verbatim}

\subsection{\JSON\ structures}\label{subsection_json}
\JSON\ is the acronym of JavaScript Object Notation, a data structure embraced
by curly parentheses and formally representing a collection of commas-separated
entries. Each entry consists of three entities: from the left, (1) the
referencing identifier i.e. the `\emph{key}', (2) followed by the
\emph{punctuator} `\texttt{:}' in the middle and (3) the `\emph{object}' on the
right:
\begin{Verbatim}[fontsize=\scriptsize,numbers=none,frame=single]
{
    <key1> : <object1>,
    ...,
    <keyN> : <objectN>
}
\end{Verbatim}
If the right object \texttt{<objectN>} is of \JSON\ datatype, the
implementation shows up in this nesting fashion:

\begin{Verbatim}[fontsize=\scriptsize,numbers=none,frame=single]
var _main_j = {
    <key1> : {
        <subkey1> : <object1>
    }
}
\end{Verbatim}

\subsubsection{Compact and dynamics construction}\label{paragraph_json_compact_construction}
The implementation of \JSON\ objects in Javascript can be achieved by
\emph{explicitly} declaring the constituents, according to the above code.
\JSON\ objects can be \emph{dynamically} populated, but not created in this
manner unless they are regarded as associative arrays. We overloaded the symbol
`\texttt{:}' to act like a binary operator that takes two input
\texttt{Array}s, including the keys and values, appearing on the left and on
the right respectively. Elements with the same index are paired to form a new
\JSON\ object. In the example below, the operator takes two input
\emph{explicit} arrays:\footnote{A list of comma-separated objects is included
between an opening and a closing square bracket.}

\begin{Verbatim}[fontsize=\scriptsize,numbers=none,frame=single]
var _a = [a,b,c]:[1,2,3];
console.log(_a);
\end{Verbatim}
Furthermore, two \emph{implicit} \texttt{Array} objects have been input
below:\footnote{An explicit array that is referred by one identifier.}

\begin{Verbatim}[fontsize=\scriptsize,numbers=none,frame=single]
var keys = [1,2,3], values = [a,b,c];
var _j = keys : values;
\end{Verbatim}
This latter declination was not implemented yet, because of the restrictions we
illustrated in the `Disclaimer' section \ref{Disclaimer}.

\subsubsection{Decorated tokens for \JSON.}\label{paragraph_json_decorated}
A nested \JSON\ structure implementation is essentially a multi--branched tree
with \emph{node distance} metric.\footnote{Defined `\texttt{immediate}' for
short.} Each node has a \emph{depth level}, which is an integer set to $0$ for
the \JSON\ object at the \emph{root} of the tree and increasing by one unit as
\JSON objects are nested in each other. The tree transversions from parent to
child nodes and back can be represented visually in the downward and upward
directions respectively. If the metric is \emph{oriented},\footnote{The
ascending and descending transversion is denoted by the positive (`\texttt{+}')
and negative (`\texttt{-}') operator respectively.} the tree transversion
returns an integer that ranks the node's depth; otherwise, if the metric is
\emph{absolute} (not--oriented), one rates the path length, i.e. the number of
steps required to traverse the tree from some start node to the destination.
The \emph{parental distance} is computed in terms of oriented metric and it
applies to the ascending direction only.

\begin{remark}
The goal is to develop tools that could improve the interaction between nested
\JSON\ objects.
\end{remark}
Let the general concept of \emph{decorated token}, an alphanumeric reserved
identifier that is formally prefixed by the symbol `\texttt{$@$}'
(\footnote{Non-conformal to the standard naming convention.}) and
conventionally playing as an alias for objects with different roles, depending
on the syntax pattern where the decoration is applied. The decorated tokens
\texttt{$@$parent} and \texttt{$@$root} have been designed to refer to \JSON\
objects of constant parental distance, that is, 1 and the node depth
respectively. Target nodes can be reached by parental direction only:
multi--branched tree transversion is not an invertible
operation,\footnote{While moving downwards to a child keeps track of the
starting node, moving upwards to the parent loses this information.} so the
exact starting child node after ascending to its parent cannot be determined,
unless the latter has one only child. Let the next example, featuring two
nested levels and the \texttt{$@$parent} decorated token:

\begin{Verbatim}[fontsize=\scriptsize,numbers=none,frame=single]
var _j1 = {
	a : 1,
	sub_j : {
    	fn1 : function(){ return @parent.a; },
    	fn2 : function(){ return @parent(2).a; }
    }
}

alert( _j.sub_j.fn1() );
\end{Verbatim}

Multiple nodes ascension is calculated iteratively using a single input number,
whose default value is 1. Then \texttt{$@$parent(2)} would \emph{tentatively}
point to the \JSON\ object at two levels up;\footnote{At present, the formal
parsing does not allow to parameterize the multiplier: for example, to input
one variable: \texttt{$@$parent(n)}.} the transpiler will throw an error,
because this decorated token does not resolve into a valid object. This is not
the case for the token \texttt{$@$parent},\footnote{Or simply
\texttt{$@$parent}.} that refers to \texttt{\_j1}. Now let

\begin{Verbatim}[fontsize=\scriptsize,numbers=none,frame=single]
var _j1 = {
    a : 1, b : 2,
    j2 : {
       j3_1 : {
            fn() : function(){ return @root.a },
            str : 's'
       },
       j3_2 : {
            fn() : function(){ return @parent.j3_1.str }
       }
    }
}

alert( _j3.fn() );
\end{Verbatim}
Regardless of the original depth level of the calling structure, the decorated
token \texttt{$@$root} goes straight, in one only instruction, to the \JSON\
object \texttt{\_j1} at top of the tree. This initial implementation not only
allows to refer to parent nodes, but also opens to composition and suggests the
possibility of quickly accessing to \emph{sibling nodes}:\footnote{Standing at
the same level and sharing the same parent node.} for example, the path
\texttt{$@$parent.j3\_1.str} in the above code.

\subsection{Reverse self--operators}\label{subsection_reverse_self_operators}
The standard self--operator syntax pattern \texttt{<id\#1><op><assignment><id\#2>}
expands as follows: \texttt{<id\#1><assignment><id\#1><op><id\#2>}.

The binary operator \texttt{<op>} is assumed to take on the two input
parameters \texttt{<id\#1>}, \texttt{<id\#2>}; it sets the result to the
variable \texttt{<id\#1>}. The direction of the operator is left--to--right,
from \texttt{id\#1} to \texttt{<id\#2>} and it finally goes back again to
\texttt{<id\#1>}.

We also have the chance of diverting the result to \texttt{<id\#2>}. This is
possible through the reverse self--operator syntax pattern. This implementation
would not make sense for sets of commutative values\footnote{Roughly speaking,
when \texttt{<id\#1><op><id\#2><assignment><id\#2><op><id\#1>}}, say the
addition operator for real numbers. So we invite the reader to consider
\texttt{String}s instead: a common, non--commutative\footnote{For the binary
operator `\texttt{+}'.} datatype. In the following example, let some
\texttt{String}s appended to each other; the reverse syntax pattern formally
swaps the operator symbols inside the old one. Here we used the notation
\texttt{=+}, which reverses the standard operator \texttt{+=} and sets the
result to the container \texttt{<id\#2>}:

\begin{Verbatim}[fontsize=\scriptsize,numbers=none,frame=single]
var _s1 = 'I', _s2 = 'am';
var _s3 = _s1;
_s3 += _s2;
@factotum.log(_s3);

_s3 = _s1; _s3 =+ _s2;
@factotum.log(_s3);
\end{Verbatim}

\subsection{The \texttt{ifchain} decisional statement}\label{subsection_ifchain}
The relationships governing the interaction of multiple conditions might be of
dependency or not. In the first case, conditions are resolved \emph{in logical
order}: a common practice in \emph{defensive programming}; they could be
assumed to belong to a chain of nested `\texttt{if}' statements. If no code is
required between the tests of the conditions, they all can be represented into
this compact version, announced by the reserved instruction `\texttt{ifchain}':

\begin{Verbatim}[fontsize=\scriptsize,numbers=none,frame=single]
ifchain ( obj_exists( _obj ), is_of_type( _obj ) )
    do_something();
else ifchain ( obj_exists( _obj2 ), has_the_property( _obj, 'prop' ) )
    do_something_else();
\end{Verbatim}
Multiple conditions are embraced between round parentheses.\footnote{This
pattern is illegal in ordinary Javascript.} For sake of clarity, the above code
has been expanded as follows:

\begin{Verbatim}[fontsize=\scriptsize,numbers=none,frame=single]
if ( obj_exists( _obj ) )
{
    if ( is_of_type( _obj ) )
    {
        do_something();
    }
}
else if ( obj_exists( _obj2 ) )
{
    if ( has_the_property( _obj, 'prop' ) )
    {
        do_something_else();
    }
}
\end{Verbatim}
This pattern gathers conditions -- normally piled up into nested blocks like
the above expansion -- into a simple list.

\subsection{Extended \texttt{switch}}\label{subsection_extended_switch}
We add the possibility of entering \POSIX\ regular expressions into the
\texttt{case} conditions of the \texttt{switch} statement, in order to test
full classes of values, rather than single elements, such as numbers or
strings:

\begin{Verbatim}[fontsize=\scriptsize,numbers=none,frame=single]
var _a = 1;
switch( _a )
{
    case /^[0-9]+$/:
        console.log( 'integer' );
    break;
    case /^[0-9]+[\.]?[0-9]+$/:
        console.log( 'decimal' );
    break;
    default:
        console.log( 'unknown' );
    break;
}
\end{Verbatim}
We used regular expressions to generate different responses, whether the
variable $\_a$ is an integer or a decimal number.

\subsection{Fork operator} The ternary operator \texttt{?} provides a compact
and comfortable syntax pattern that squeezes the \texttt{if} and can be
implemented within larger expressions. In any case, the return value can only
be set to one variable per call. The `fork' operator \texttt{|<} aims at
overcoming such a structural restriction. The basic pattern is very similar to
the ternary version, in which the colon operator `\texttt{:}' separates the
tokens, but it accepts three parameters:

{\centering \texttt{<condition> |< <id\#1> : <id\#2> : <value> } }

The target container for \texttt{<value>} would be \texttt{id\#1} or
\texttt{id\#2} if the \texttt{<condition>} holds or not respectively:

\begin{Verbatim}[fontsize=\scriptsize,numbers=none,frame=single]
var a = 0, b = 0, c = 3;

c > 2 |< a : b : 4;

console.log( "c", c, "a", a, "b", b );
\end{Verbatim}
The value 4 will be expected to be set to a, whereas \texttt{b} will be left to
the original value 0. The \texttt{fork} operator could also accept four
parameters at most:

{\centering \texttt{<condition> |< <id\#1> : <id\#2> : <value\#1> : <value\#2>
} }

\noindent In practice, let the code

\begin{Verbatim}[fontsize=\scriptsize,numbers=none,frame=single]
var a = 0, b = 0, c = 3;

c < 2 |< a : b : 4 : 5;
console.log( "c", c, "a", a, "b", b );
\end{Verbatim}
The first two are destination variable identifiers (\texttt{a}, \texttt{b})
again, but the last couple (\texttt{4}, \texttt{5}) to the objects that will be
set to the variables \emph{respectively}: then \texttt{a=4} if \texttt{c < 2}
holds, and \texttt{a=5} otherwise.

\subsection{Binary operators}
\OBJS\ implements the following binary operators for the evaluation of values
and the return of the left or right parameter whether the far left condition
holds or not respectively, i.e. according to the following pattern:

\begin{Verbatim}[fontsize=\scriptsize,numbers=none,frame=single]
\texttt{<return-value><assignment><left-par><op><right-par>}
\end{Verbatim}

\noindent or, concisely.

\begin{Verbatim}[fontsize=\scriptsize,numbers=none,frame=single]
\texttt{var a = b <op> c}
\end{Verbatim}

\begin{tabular}{|p{0.15\linewidth}|p{0.25\linewidth}|p{0.4\linewidth}|}
    \hline
    \emph{Operator} & \emph{Definition} & \emph{Condition}\\
    \hline
    \footnotesize\texttt{??} & Null coalescing & \texttt{b} is not \texttt{null}\\
    \hline
    \footnotesize\texttt{?:} & Elvis operator & \texttt{b} is not false or 0\\
    \hline
    \footnotesize\texttt{?:} & Safe navigation &\texttt{b} is \texttt{null} \\
    \hline
    \footnotesize\texttt{?==} & loose equality & \texttt{b} and \texttt{c} are equal in value only\\
    \hline
    \footnotesize\texttt{?===} & strict equality & \texttt{b} and \texttt{c} are equal in value and type\\
    \hline
    \footnotesize\texttt{?<} & lesser & \texttt{b < c}\\
    \hline
    \footnotesize\texttt{?>} & greater & \texttt{b < c}\\
    \hline
    \footnotesize\texttt{?<=} & lesser and equal & \texttt{b <= c}\\
    \hline
    \footnotesize\texttt{?>=} & greater and equal & \texttt{b >= c}\\
    \hline
\end{tabular}

\subsection{Decorated identifiers} The entries in the table below are not part of
Javascript standards. Decorated identifiers are implemented for \OBJS\ internal
usage exclusively. They are prefixed by the symbol `texttt{\@}' and applied in
different contexts; as long as we need them, we will run into their
details.\vspace{0.3cm}

\begin{tabular}{|p{0.15\linewidth}|p{0.65\linewidth}|}
\hline
Identifier & Purpose\\
\hline
\texttt{\@1, \@2, \dots} & reference to arguments inside \texttt{\#overload} headers \\
\hline
\texttt{\@arg} & applied inside the `argument safe' syntax pattern\\
\hline
\texttt{\@counter} & applied to some special syntax patterns\\
\hline
\texttt{\@file} & retrieves the file name where the running code belongs to\\
\hline
\texttt{\@line}, \texttt{\@column} & coordinates of this token inside the text code\\
\hline
\texttt{\@namespace} & retrieves the namespace in use\\
\hline
\texttt{\@parent}, \texttt{\@root} & applied to nested JSON structures\\
\hline
\texttt{\@src} & reference to source object inside \texttt{\#overload typecasting} header\\
\hline
\end{tabular}

\subsection{Namespaces}
Inspired by accidents that may occur during the development of large back--end
projects in \PHP, we implemented \emph{namespaces} for Javascript, because
front--end services are also becoming increasingly large and
demanding.\footnote{The \texttt{use strict} directive in Javascript prevents to
refer to undeclared variables. \PHP\ and Javascript do not feature a rigid liveness
analysis, for tracking multiple declarations of the same variable within the same
scope for example.}

Namespaces are logical domains, where identifiers are re--organized under specific
naming conventions, to prevent collisions between homonymous objects, possibly
declared elsewhere. That convention shows close syntax similarities with the path
pattern of filesystems,\footnote{Particularly, of Unix--like systems.} where unique
referrers are also required.

\begin{Verbatim}[fontsize=\scriptsize,numbers=none,frame=single]
namespace ns1
var _a1 = 1;
exit namespace

var _f = 1 + 2 / 3;
alert( _f );
alert( ns1\_a1 );
\end{Verbatim}

When isolated, the reserved word \texttt{namespace} automatically closes the
previous domain, if any, and it opens a new one. On the contrary, \texttt{exit
namespace} forcingly closes the current domain:

\begin{Verbatim}[fontsize=\scriptsize,numbers=none,frame=single]
namespace ns1
var _a1 = 1;

namespace ns2
var _a1 = 2;

exit namespace

alert( ns1\_a1 );
alert( ns2\_a1 );
\end{Verbatim}

Namespace domains can be embraced by curly parentheses, for sake of clarity about
their contents:

\begin{Verbatim}[fontsize=\scriptsize,numbers=none,frame=single]
namespace ns1 {
    var _a = 1, _b = 2;
    function fn(x){ return x + 1; }
}

namespace ns2 {
    var _a = 3, _b = 4;
}

var _c = 0; //everything out does not belong to namespaces
console.log( ns1\_a, ns1\_b );
console.log( ns2\_a, ns2\_b, _c );
\end{Verbatim}

If namespace paths are too long, the reserved term `\texttt{as}' may help to set up
shorter aliases:\footnote{There exists a minor risk of falling back again into
collisions between aliases anyway.}

\begin{Verbatim}[fontsize=\scriptsize,numbers=none,frame=single]
namespace \lev1\lev2\lev3;
var _a = 1;

use \lev1\lev2\lev3 as short3;
console.log( short3\_a );
\end{Verbatim}

Namespaces could also include \texttt{class} declarations:

\begin{Verbatim}[fontsize=\scriptsize,numbers=none,frame=single]
namespace \lev1\lev2\lev3
class cls{
    constructor(){ /* do nothing */ }
}
exit namespace

var _c = new \lev1\lev2\lev3\cls;
\end{Verbatim}

\section{Data typification}
\subsection{Typified declarations}\label{subsection_typified_declaration}
In order to introduce this feature and to compare this new syntax pattern with
the standard Javascript implementations, we first reported some code samples,
paired in lines, where each new syntax pattern, including abstract
\texttt{<tokens>}, precedes the related ordinary version:

\begin{Verbatim}[fontsize=\scriptsize,numbers=none,frame=single]
<datatype> <referrer>; // declaration without arguments
    var <referrer> = new <datatype>();

<datatype> <referrer> = _arg; // single argument
    var <referrer> = new <datatype>(_arg);

// multiple declarations
<datatype> <referrer#1>, <referrer#2>;
    var <referrer#1> = new <datatype>(), <referrer#2> = new <datatype>();

// multiple arguments
<datatype> <referrer> = (_arg1,_arg2);
    var <referrer> = new <datatype>(_arg1,_arg2);
\end{Verbatim}
When there are more than one parameter, they are embraced between round
parentheses:

\begin{Verbatim}[fontsize=\scriptsize,numbers=none,frame=single]
complex _imag = ( 0, 1 );
\end{Verbatim}

This typified declaration shows flexibility by supporting nested constructions
and multiple declarations:\footnote{In order to let this pattern deal with the
standard \texttt{var}/\texttt{let} declaration, we followed this policy: (1) to
use \texttt{let} only inside function and statement bodies and (2) \texttt{var}
outside, that is, in the main code thread.}

\begin{Verbatim}[fontsize=\scriptsize,numbers=none,frame=single]
complex _a = ( int _real = 1, int _imag = 0 ), _b = ( complex (1,2) );
\end{Verbatim}

In the next example, we declared an object of \texttt{complex} datatype and we
multiplied it by 2, and we finally added the resulting product to 1:

\begin{Verbatim}[fontsize=\scriptsize,numbers=none,frame=single]
complex _a = 1.1;
@factotum.alert( _a * 2 + 1 );
\end{Verbatim}

Although it is redundant in terms of operators precedence, we will edit the
formula once more, this time putting the round parentheses around the product,
\texttt{(\_a*2)+1}, to make computations more understandable: the above formula
first multiplies a \texttt{complex} object by a \texttt{Number} object;
finally, the returned \texttt{complex} product adds to a \texttt{Number}. The
object \texttt{\_a} is of non--standard \texttt{complex} datatype, so
\texttt{\_a*2+1} requires the translation into this chain of method calls:

\begin{Verbatim}[fontsize=\scriptsize,numbers=none,frame=single]
complex _a = 1.1;
@factotum.alert(_a.mul(2).add(1));
\end{Verbatim}

As we change the above formula into \texttt{\_a+1*2}, where 1 and 2 are
(standard) \texttt{Number} objects,\footnote{The star operator `\texttt{*}'
follows the default implementation and it has not need to be overloaded.} a new
translation will be required:

\begin{Verbatim}[fontsize=\scriptsize,numbers=none,frame=single]
complex _a = 1.1;
@factotum.alert( _a.add( 1 * 2 ) );
\end{Verbatim}

\begin{remark}
The operators overloading could represent an efficacious alternative to method
calls, being impracticable to work out formulas with increasing level of
complication.
\end{remark}

\subsection{Functions in Javascript}\label{subsection_functions}
Functions are \emph{finite, sequential aggregations of statements}, assembled
in accordance with the traditional I/O model: when input objects (the
arguments) are fed into the function, it may optionally \texttt{return} one
value or not, before it terminating its run.\footnote{Javascript treats
functions as datatypes, with the ability of instantiating them as new objects
with members and \texttt{prototype}d methods.} The function represents a
facilitation to run several commands after only one call; it could set up a
\emph{one--to--one} or \emph{one--to--many} relation from the (one) call to
(one or many) statements therein. Functions are of \emph{modular nature}, as
their design is based on a single block of reusable code that is separated from
the main code thread.

\subsubsection{A short synopsis on the declaration pattern}
This is the abstract, formal pattern to declare functions in Javascript:
\begin{Verbatim}[fontsize=\scriptsize,numbers=none,frame=single]
<function-header>
{
    <function-body>
}
\end{Verbatim}

The abstract \texttt{<function-header>} expands into the \texttt{header} at the
top and the following \texttt{body} of statements:
\begin{Verbatim}[fontsize=\scriptsize,numbers=none,frame=single]
function <name>( <arg-id#1>, ... )
{
    // all statements are optionals: body could be empty
    <statement#1>
    <statement#2>
    ...
    <return-statement>
}
\end{Verbatim}
The token \texttt{<name>} refers to the function identifier, for the next
calls.\footnote{This token is not mandatory: if omitted, functions are assumed
as \emph{anonymous}. But just apparently at users scope, as the Javascript
engine would apply a referrer to record the function into memory at a deeper
level.} Function calls execute statement(s) within its body in the same order
as they appear throughout the code; the last statement could optionally include
the reserved word \texttt{<return>}, in order to send an object to the main
code thread.\footnote{The \texttt{<return>} token is not mandatory. Given the
\texttt{return} and the \texttt{object} token inside the function body, three
cases may occur: (1) both tokens are given, (\texttt{return object}): the
function sends the variable \texttt{object} to the calling code; (2) only the
\texttt{return} token (\texttt{return ;}) which actually returns no object, (3)
none of the two tokens, in order to not perform any return action at all.}
Therefore it makes sense to implement datatype specification into function
declaration.

\subsubsection{Typified function declaration}
As we did for typified variables, we are going to typify function declaration
too: both have similar declarative patterns. We add two new specifications,
inherent to the new datatypes mentioned and collectively classified as
\emph{function typification}. The abstract token \texttt{<return-datatype>}
refers to the datatype returned by the function call, whereas
\texttt{<arg-datatype\#n>} points to the datatype of each input argument
\texttt{<arg-id\#n>} between round parentheses:

\begin{Verbatim}[fontsize=\scriptsize,numbers=none,frame=single]
<return-datatype> function <name>( <arg-datatype#1><arg-id#1>, ... )
{
    <function-body>
}
\end{Verbatim}

This pattern can be implemented in \emph{weak} or in the \emph{strong} version.
We first present two examples of \emph{weak function typification}. The weak
character is due to the omission of the \texttt{<return-datatype>}, which is
assumed as a generic \texttt{Object} by default. We declared below two
homonymous functions `\texttt{fn}' featuring different typifications: the first
takes only one input parameter of \texttt{complex} datatype, whereas the second
declaration takes only one \texttt{Number} datatype

\begin{Verbatim}[fontsize=\scriptsize,numbers=none,frame=single]
function fn(complex a){ return "complex"; }
function fn(Number a){ return "number"; }

console.log( fn( 1 ) );
console.log( fn( new complex(1,2) ) );
console.log( fn( "String") ) );
\end{Verbatim}
All declarations of homonymous objects are overridden by the last in line.
\OBJS\ uses the arguments datatype to recognize the exact typified declaration
and then to mangle it. The \emph{mangling technique} allows the registration of
multiple typified declarations of a same homonymous function, in order to find
the exact match for the requested typification. A warning would be thrown if
the call finds no match at run-time. For example, the simple call
\texttt{fn("String")} at row 6 would throw a warning because no function that
takes one only \texttt{String} argument has been registered. This advanced
example combines weak typification and formulas:

\begin{Verbatim}[fontsize=\scriptsize,numbers=none,frame=single]
function fn(complex c){ return c; }

complex c1 = (1,2);
var _sum = c1 + fn( new complex(1,2) );
\end{Verbatim}
In the economy of this transpiler, the weak typification breaks the chain of
formal datatype recognition: while the left addend \texttt{c1} is correctly
detected as of \texttt{complex} datatype, the addend on the right, resulting
from this preventive extraction

\begin{Verbatim}[fontsize=\scriptsize,numbers=none,frame=single]
var _o_1 = fn( new complex(1,2) );
var _sum = c1 + _o_1;
\end{Verbatim}
is assumed as of generic \texttt{Object} datatype. The generic datatype can be
viewed as a failure along the datatype recognition process, like an
`indetermination status' propagating to subsequent values that depend on the
call to \texttt{fn}; thus the variable \texttt{\_sum} would be a
\texttt{Object} too in the cascade fashion. The solution is to use the
`\emph{strong function typification}', a more performant pattern that includes
the \texttt{return-datatype} in the function \texttt{header} (\footnote{The
\texttt{return} is mandatory here.}) on the left of the \texttt{function}
token:
\begin{Verbatim}[fontsize=\scriptsize,numbers=none,frame=single]
complex function fn(complex c){ return c; }

complex c1 = (1,2);
var _sum = c1 + fn( new complex(1,2) );
\end{Verbatim}
The formal datatype recognition chain cannot be broken now. The token
\texttt{<return-datatype>} fills the gap left by the weaker pattern and serves
the same purpose as in the typified declaration. Thus the variable
\texttt{\_sum} includes the value resulting from the action induced by the
operator `\texttt{+}', which takes two \texttt{complex} objects.

\begin{remark}
Weak and strong function typifications coexist for granting full compatibility
between standard and \OBJS\ patterns.
\end{remark}

\noindent Both typifications can be extended to the \emph{arrow function}
pattern:

\begin{Verbatim}[fontsize=\scriptsize,numbers=none,frame=single]
//weak typification
var fn = (complex a, complex b) => a * b;

//strong typification
complex fn = (complex a, complex b) => a * b;
\end{Verbatim}

With the necessary adaptations, the strong typification can be exported from
functions to methods within \JSON\ structures:

\begin{Verbatim}[fontsize=\scriptsize,numbers=none,frame=single]
var _j = {
    Number fn : function( Number a ){ return a; },
    complex fn : function( complex a ){ return a; }
}

var _c = _j.fn(new complex(1,2));
var _doubled = _c * 2;
@factotum.alert( _doubled );
\end{Verbatim}

\noindent Or to the declarations of \texttt{prototype}d members:

\begin{Verbatim}[fontsize=\scriptsize,numbers=none,frame=single]
//Javascript functions are also intended as datatypes
function obj(){}

obj.prototype.method = complex function( complex c ){
  console.log( "complex", c );
  return c;
}

obj.prototype.method = function( o ){
  console.log( "object", o );
  return o;
}

var _obj = new obj();
_obj.method( new complex(1,2) );
_obj.method( 1 );
\end{Verbatim}

\noindent As well as to \texttt{class} methods:

\begin{Verbatim}[fontsize=\scriptsize,numbers=none,frame=single]
class cls{
	constructor(){}
    complex fn(complex c){return c;}
    Number fn(Number n){return n;}
    //String fn(String str){return str;}
}

var _obj = new cls();
var _n = _obj.fn( 1 );
var _c = _obj.fn( new complex(1,2) );
var _str = _obj.fn( "a" );
@factotum.alert(_c); @factotum.alert(_n);
@factotum.alert(_str);
\end{Verbatim}

The above code throws a warning: the \texttt{String} typified version of the
`\texttt{fn}' method was called, but not declared yet (here commented). As line
5 is uncommented, the newly compiled code will be able to find the exact match.

\subsubsection{Safe argument}
The purpose of this syntax pattern is to ensure that function arguments are of
the specified datatype. If no such match is found, the argument is typecasted
to the class itself (provided that the conversion has been encoded). This
pattern applies to input arguments inside the function header only. The
argument identifier must be prefixed by the symbol `\texttt{\@}' and preceded
by the datatype identifier. It applies to weak and strong typification of
functions.
\begin{Verbatim}[fontsize=\scriptsize,numbers=none,frame=single]
function fn( complex @arg ) { return arg; }
\end{Verbatim}

\section{Syntactic sugars}
We reported below two `syntactic sugars' that may help avoid long codes.
\subsection{Block repeater}
This simple pattern could avoid long \texttt{for} loop headers, as it
`multiplies' the block of instructions:

\begin{Verbatim}[fontsize=\scriptsize,numbers=none,frame=single]
4 * {
	console.log( @counter );
}
\end{Verbatim}
This special environment re--parses into the ordinary \texttt{for}--loop. The
internal decorated identifier \texttt{\@counter} works to build up the ordinary
code and keeps track of the loop counter increment.

\subsection{Ordered sequences.} This syntax pattern returns an array of values
ranging from end to end immediately:

\begin{Verbatim}[fontsize=\scriptsize,numbers=none,frame=single]
var _a = (1,...,10);
console(_a);
\end{Verbatim}
\noindent This pattern can be applied to alphabetic sequences:

\begin{Verbatim}[fontsize=\scriptsize,numbers=none,frame=single]
var _a = ("a",...,"z");
console(_a);
\end{Verbatim}
\noindent Similarly to \Cpp, it is possibile to simulate the call-by-reference
syntax, as follows:

\begin{Verbatim}[fontsize=\scriptsize,numbers=none,frame=single]
function fn( Number & v ){ v++; }

var a = 1;
fn( a );
console.log( a );
\end{Verbatim}
\noindent The value of variable \texttt{a} is expected to store the value
\texttt{2}.

\section{Optimizations}
We browse some optimization strategies to lessen the code; they are elaborated
when the transpiler \texttt{pragma optimize} mode is set on.

\subsection{Unreferenced function argument}
The next example shows a function declaration including two parameters
\texttt{a} and \texttt{b}; only \texttt{a} is referenced within the body, so
\texttt{b} can be safely removed and lessen the arguments stack.

\begin{Verbatim}[fontsize=\scriptsize,numbers=none,frame=single]
function fn( complex a, complex b ) {
	return a;
}
\end{Verbatim}

\subsection{Unreachable or dead code}
We provided some examples including code which will not be run for different
reasons. Below, no line of code will be run after the \texttt{return}
statement; thus it can be dropped out:

\begin{Verbatim}[fontsize=\scriptsize,numbers=none,frame=single]
function fn( complex a, complex b ) {
	return a+b;
    a /= b;
}
\end{Verbatim}
Blocks of code subdued to the evaluation of condition statement will not be run
if the latter evaluates to \texttt{0}, \texttt{false}, \texttt{null} or
\texttt{undefined}; here again, the (block of) code can be safely dropped out:

\begin{Verbatim}[fontsize=\scriptsize,numbers=none,frame=single]
var a = 1, b = 0 ;

if ( a && b ) { doSomething(); }
\end{Verbatim}

\subsection{Isolated deletion}
Those identifiers that are only referenced will be assumed to be irrelevant
throughout the run and will be removed:

\begin{Verbatim}[fontsize=\scriptsize,numbers=none,frame=single]
function fn( complex a, complex b ) {
	return a;
}
\end{Verbatim}

\subsection{Constant folding}
The result of arithmetic expression, including constant values only, will be
automatically calculated and will replace the formula:

\begin{Verbatim}[fontsize=\scriptsize,numbers=none,frame=single]
var _a = 1 + 2 / 4;
//this line is replaced by : var _a = 1.5;
\end{Verbatim}

\subsection{Constant propagation}
The constant \texttt{1}, assigned to \texttt{\_a} is `propagated', i.e. the
identifier is replaced with the same value:

\begin{Verbatim}[fontsize=\scriptsize,numbers=none,frame=single]
var _a = 1;
var _b = _a;

/* the code is optimized to
var _a = 1;
var _b = 1;
*/
\end{Verbatim}
Constant folding and propagation techniques can be jointly applied, like in the
case below, where the numerical expression is elaborated and the result is
propagated to the subsequent occurrences of \texttt{\_a}:

\begin{Verbatim}[fontsize=\scriptsize,numbers=none,frame=single]
var _a = 1 + 5;
var _b = _a ;

/* the code is optimized to
var _a = 1;
var _b = 1;
*/
\end{Verbatim}

\section{Preprocessors}
\subsection{The \texttt{\#overload} directive.}\label{subsection_overload_predirective}
Overloading represents \emph{the ability of customizing syntax patterns by
binding them to some block of code}. Developer should not read such a
`personalization' as a threat for the concepts defragmentation, or as a
super--structure, like a new language within the official language:

\begin{remark}
Overloading does not pertain to new procedures because it is a portable
artifice for coding the same actions more efficiently by replacing them with
easier-to-read patterns.
\end{remark}
Like the function declaration, the overloading process consists of three
stages: ($a$) \emph{registration} of settings (like declaration) and ($b$)
\emph{recognition} of the overloaded pattern: the operator and the arguments
datatypes (like the function call); ($c$) the \emph{application} to ordinary
code. Overloading must be declared outside the compound statements (function
declarations, loops, decisions, \dots), and preferably before the main code
thread.

\OBJS\ implements the multi--purposed \texttt{\#overload} directive that
includes some patterns depending on the candidate object to be overloaded:
possibly, an (1) operator, (2) function, (3) reserved word, (4) event, (5)
typecasting.\footnote{Overloading is part of so called \emph{extensible
languages}, which include tools to modify the semantics of patterns and
operators \cite[p. 58]{AhoUllman-1972}.}

\subsubsection{Symbolic operators}
Let a `\emph{symbolic operator}' be any typeset character which is not
alphanumeric or one of these punctuators: comma, full stop, parentheses,
quotes, semicolon and tract. Operator overloading relies on datatype
typification (\S \ref{subsection_typified_declaration}) and it can be properly
listed among the translation activities, strictly speaking. Overloading applies
to symbolic operators of unary or binary family only, that is, those taking one
or two input arguments. Like the function declaration, the abstract syntax
pattern consists of parameters and settings in the top \texttt{header} and of a
\texttt{body} of instructions:

\begin{Verbatim}[fontsize=\scriptsize,numbers=none,frame=single]
#overload <associativity> <self> <priority> <role>
    <return-datatype> <symbolic-operator>
    ( <input-arguments> )
{
    <body-with-statements>
}
\end{Verbatim}

The variety of parameters in this \texttt{header} offers a sufficient degree of
freedom to customize the operator behavior. The basic set of mandatory
parameters includes the tokens \texttt{<return-datatype>} and
\texttt{<symbol>}. Other parameters are apparently optional as they are not
required \emph{if overloading does not need to be sharply customized}. For
instance, the parameter \texttt{<self>} would be optional by default, but it is
required in order to register self--operators, such as
`\texttt{+=}'.\footnote{Or also `\texttt{-=}' for the reversed,
non--commutative version. Self--operators are assumed to be binary and to
`compress' the usual formulation: the return value is assigned to the same left
argument.} The parameter \texttt{<role>} is optional, but required to set up
custom precedence during the evaluation of expressions; operators precedence is
just a convention that we may change arbitrarily, with regard to the datatypes
involved. When optional parameters are omitted, the transpiler attempts to
recognize them, according to default settings and to the operator symbols.

We are now able to overload two versions of the same symbolic operator, now
endowed with distinct associativity: either left--to--right (LR,
`\texttt{prefix}') or right--to--left (RL, `\texttt{postfix}'). Then we
overloaded twice this unary operator `\texttt{!}' to take an argument of
\texttt{complex} datatype:\footnote{The identifier \texttt{$@$factotum} refers
to an object included in the \OBJS\ environment.}

\begin{Verbatim}[fontsize=\scriptsize,numbers=none,frame=single]
#overload prefix operator complex !(complex @1){ return new complex(0,
@1.imag); }

#overload postfix operator complex !(complex @1){ return new complex(@1.real,
0); }

complex _z = (1,1); // new feature: typified declaration

@factotum.alert(_z!); @factotum.alert(!_z);
\end{Verbatim}
The semantics of the symbol `\texttt{!}' have not been associated to the usual
\texttt{logical not} operator: \emph{operators overloading is not restricted to
preset semantics and symbolic definitions}. In the next example, we picked up a
sequence of arbitrary characters `\texttt{!!!}' and registered it as a custom
binary operator that takes two input integers (\footnote{Encoded by the
decorated tokens `\texttt{$@$1}' and `\texttt{$@$2}'.}) and returning an
\texttt{Array} object, filled by the ordered sequence of values ranging between
the ones set into the two decorated parameters:

\begin{Verbatim}[fontsize=\scriptsize,numbers=none,frame=single]
#overload operator Array !!! (Number @1, Number @2) {
    let _a = [];
    for( let _i = @1; _i <= @2; _i++ )
        _a.push(_i);
    return _a;
}

alert( 1 !!! 10 );
\end{Verbatim}
\OBJS\ can also to customize `\emph{polyadic}' operators which are not symbolic
and take a larger number of input arguments.

\begin{Verbatim}[fontsize=\scriptsize,numbers=none,frame=single]
#overload polyadic Boolean (Number @1) among (Number @2, Number @3) { return
(@2 <= @1) && (@1 <= @3); } alert( 5 among( 1, 2 ) );
\end{Verbatim}

\subsubsection{Functions}
Function identifiers can be overloaded by different comma--separated aliases.
In the next example, the method \texttt{tg}, belonging to the custom class
\texttt{complex} and evaluating the trigonometric tangent function (accepting
one input value of \texttt{complex} datatype), has been overloaded for being
called through \texttt{tan} and \texttt{tang}. In opposite to the present
example, this overloading could be useful to differ homonymous methods that
belong to different class definitions. Notice that it is essential to input
argument datatype(s), in order to recognize the original function call pattern.

\begin{Verbatim}[fontsize=\scriptsize,numbers=none,frame=single]
#overload function complex tg alias tanX, tangX(complex @1){
    return @1.tg();
}

complex _z = (1,0);

var _t1 = tg(_z); // custom defined to work with complexes

var _t2 = tanX(_z); var _t3 = tangX(_z);

@factotum.alert( _t1 + _t2 + _t3 );
\end{Verbatim}

The function \texttt{tg} has been already registered in the default settings
and the above overload will just add the new aliases \texttt{}

\subsubsection{Commands}
The overloading of `\texttt{command}' reworks the essentials of the operator
concept, as it is transposed to the naming convention acknowledging alphabetic
characters only. The first argument datatype was set to `\texttt{generic}', a
fictional datatype meant as an \OBJS coding convention to accept input
parameters of any given datatype, for the overload recognition pattern to work
in any circumstance.

\begin{Verbatim}[fontsize=\scriptsize,numbers=none,frame=single]
#overload command boolean is(generic @1, String @2) { return RegExp( @2, "i"
).test( typeof @1 ); }

console.log( 2 is "complex" ); console.log( "hello" is "string" );
\end{Verbatim}
In the next similar case, we overloaded a command that check the existence of
an element inside the input array.

\begin{Verbatim}[fontsize=\scriptsize,numbers=none,frame=single]
#overload command boolean inside(generic @1, Array @2) { return
@2.includes(@1); }

console.log( 1 inside [1,2,3] );
\end{Verbatim}

\subsubsection{Overloading the reserved: speaking in Javascript} This side
feature customizes the official reserved words in order to have a version
closer to human language: it could be beneficial to beginners (especially,
younger ones), or to people liking Javascript code to be declined into their
native languages.

\begin{Verbatim}[fontsize=\scriptsize,numbers=none,frame=single]
//set language recognition to FR(French) only
#overload reserved LANG FR alors as DROPPABLE #overload reserved LANG FR si as
if #overload reserved LANG FR est as === #pragma translator FR

var _a = 1; si( _a est 1 ) alors alert( 'Bonjour' );
\end{Verbatim}
\noindent which pre--compiles into the ordinary code:

\begin{Verbatim}[fontsize=\scriptsize,numbers=none,frame=single]
var _a = 1; if ( _a === 1 ) alert ( 'Bonjour' );
\end{Verbatim}
The French term `alors' (`then' in English) is redundant and not featured among
the official Javascript specifics, so it will be dropped (set as
\texttt{droppable}). The standard ISO 3166--1 alpha--2 convention was adopted
for the language two--letters identifier. Strictly speaking, this is not about
`overloading' now, as this implementation more precisely concerns of
`replacement'. We continued to use the same preprocessor for sake of
simplicity.

This feature could be useful to customize the reserved words, for the code to
look easy to be read, especially for practitioners or anyone else liking to
read code in one's own native language. Languages are referred according to the
ISO 3166--1 alpha--2 coding convention.\vspace{0.5cm}

\begin{Verbatim}[fontsize=\scriptsize,numbers=none,frame=single]
#overload reserved LANG IT allora as DROPPABLE #overload reserved LANG IT é as
=== #overload reserved LANG IT se as if #pragma translator IT //set language
recognition to IT only

var _a = 1; se( _a é 1 ) allora alert( 'ciao' );
\end{Verbatim}
\noindent which pre--compiles as follows:

\begin{Verbatim}[fontsize=\scriptsize,frame=single]
var _a = 1; if ( _a === 1 ) {
    alert ( 'ciao' ) ;
}
\end{Verbatim}

\subsubsection{Code event response: overloading the
events}\label{subsection_code_event_response} Inspired by liveness analysis, we
see code as a \emph{living entity}. Not evidently endowed of self--conscience,
but of lifetime and being able to get inputs and to react to them: code is not
just as a collection of actions, but something having its story of events that
flow one after the other. Events can be tracked down and overloaded, like we
did for operators. Specific code is triggered and run in response to the input
events: for example, when a new element is pushed into an array, or a function
is called.

\subsubsection{Events timeline.} Event overloading can be registered for
responses that are tracked \emph{before} or \emph{after} the events. The
pattern \texttt{on\_before\_<event>} is called immediately before the event,
whereas \texttt{on\_<event>} runs after the event has occurred. For example,
let the conventional term `\texttt{decl}' refers to object declarations; the
compositions `\texttt{on\_before\_decl}' and `\texttt{on\_decl}' bind to two
events triggered before and after the declaration respectively. We overloaded
these two events for the reserved token `\texttt{var}'.\footnote{Alternatively,
it could be applied to `\texttt{const}' or to `\texttt{let}' declaration.}

\begin{Verbatim}[fontsize=\scriptsize,numbers=none,frame=single]
#overload event on_before_decl to var
    { console.log('before_declaration'); }
#overload event on_decl to var
    { console.log('after_declaration'); }

var _a = 1;
\end{Verbatim}
Events overloading could be subjected to restrictions or not, according to the
token value \texttt{<target-object>}:

\begin{Verbatim}[fontsize=\scriptsize,numbers=none,frame=single]
#overload event <events-name> to <target-object>
    ( <input-arguments> )
{
    <body-with-statements>
}
\end{Verbatim}
The token \texttt{<target-object>} allows to register the event and to bind it
to one or to multiple objects. For the registration of the family of events
\texttt{decl}, the target could be a variable identifier or reserved words open
to declaration: `\texttt{const}', `\texttt{let}', `\texttt{var}'.

This is the full list of overloadable events: \emph{assignment},
\emph{declaration}, \emph{object instantiation} and \emph{destruction},
\emph{function} and \emph{method call}, \emph{pushing into} and \emph{popping
from} one array.\footnote{The abstract token \texttt{<event>} would be encoded
by \texttt{assign}, \texttt{decl}, \texttt{new}, \texttt{delete},
\texttt{function\_call}, \texttt{method\_call}, \texttt{array\_push} and
\texttt{array\_pop}.}

\subsubsection{Array popping and pushing.}\label{subsection_event_array}
If elements are \emph{pushed} into or \emph{popped} from an array, the events
\texttt{on\_array\_push} and \texttt{on\_array\_pop} will be triggered
respectively. The array plays as the \texttt{<target-object>} required in the
event registration. Here the digit 1, in the \emph{decorated} variable
`\texttt{$@$1}', runs like an index to the first element of the 1--index based
array of the target objects, announced by the reserved token \texttt{<to>}:
thus `\texttt{$@$1}' stands for `\texttt{\_a}' and `\texttt{$@$2}' for
`\texttt{\_b}'.

\begin{Verbatim}[fontsize=\scriptsize,numbers=none,frame=single]
#overload event on_array_push to _a, _b
    { console.log(@1.length, @2.length); }

var _a = [];
for( var _i = 0; _i < 10; _i++ ) _a.push(1);

_b.push('string');
console.log( _a, _b );
\end{Verbatim}

\subsubsection{Declaration, assignment and multiple
events.}\label{subsection_event_assignment} The events `\texttt{on\_decl}'
(object declaration) and `\texttt{on\_assign}' (object assignment) have been
overloaded for variables \texttt{a} and \texttt{b}:

\begin{Verbatim}[fontsize=\scriptsize,numbers=none,frame=single]
#overload event on_decl, on_assign to a, b
    { console.log('response'); }

var a = 1, b = 2;
    a = 2, c = 4;
\end{Verbatim}

\noindent All declarations and assignments can be affected:

\begin{Verbatim}[fontsize=\scriptsize,numbers=none,frame=single]
#overload event on_decl, on_assign to @all
    { console.log('triggered'); }

var a = 1, b = 2; // declarations
    a = 2; // assignment
\end{Verbatim}
When the token  \texttt{<target-object>} in the declaration takes on the
decorated value `\texttt{$@$all}', the event registration extends to any
object: in the code below, the string \texttt{`triggered'} will be displayed
when any object has been declared or when one value has been assigned:

\subsubsection{Object instantiation.}\label{subsection_event_new} The event
`\texttt{on\_new}' is triggered when objects have been instantiated. The
overload \texttt{header} includes the reserved token `\texttt{to}': this
specification announces that this event is triggered if the
\texttt{<target-object>} is of \texttt{complex} datatype only:

\begin{Verbatim}[fontsize=\scriptsize,numbers=none,frame=single]
#overload event on_new to complex
    { console.log('instantiation'); }

// typified declaration
complex _c1 = (1,2);
// standard declaration and instantiation
var _c2 = new complex(3,4);

var _s = _c1 + _c2;
var _n = new Number(1);
\end{Verbatim}
This code indicates that only two events \texttt{on\_new} have been triggered:
one after the explicit typified declaration \texttt{complex \_c1} and the
second after the standard instantiation \texttt{\_c2 = new complex(3,4)}. No
\texttt{on\_new} event is triggered for the instantiation of the
\texttt{Number} object because this datatype has not been mentioned in the
event \texttt{<header>}.\footnote{But we could fix it by applying this code:
\texttt{\#overload event on\_new to complex, Number}.}

\subsection{Typecasting}\label{subsection_typecasting}
\OBJS\ performs some liveness analysis to determine lifetime and properties of
variables, as well sd \emph{typecasting}, defined as the \emph{creation}
(`casting') of a new object of class \texttt{A} \emph{from} another object of
class \texttt{B}. This process can be thought as a \texttt{datatype
conversion}, carried out in two stages: (1) the \emph{registration} of the
handler for the conversion from source to destination datatype, (2) the
\emph{parsing} of the typecasting syntax pattern, in order to run the handler
and activate the conversion, assuming that stage 1 has covered the input
datatype.

Stage 1 can be completed using the traditional method \texttt{typecasting},
which can be implemented as a \texttt{prototype} or as a \texttt{class}
member.\footnote{Depending on the nature of the container: \texttt{function} or
\texttt{class} respectively.} Otherwise, typecasting can be registered as an
overloading using this abstract syntax pattern:
\begin{Verbatim}[fontsize=\scriptsize,numbers=left,frame=single]
#overload typecasting <source-datatype> to <destination-datatype>
{
    <body-of-statements>
}
\end{Verbatim}

\noindent which manifests in this example:
\begin{Verbatim}[fontsize=\scriptsize,numbers=left,frame=single]
#overload typecasting complex to segment
{
	return new segment( 0, 0, @src.real, @src.imag );
}
\end{Verbatim}
The token \texttt{$@$src} is defined as \emph{decorated token}, a standard
feature in the \OBJS\ environment: it is a class of internal objects, prefixed
by `$@$' and performing a variety of tasks depending on the context in which
they appear. Regarding stage 2, the typecasting can appear \emph{semantically}
similar to the typified declaration (subsection
\ref{subsection_typified_declaration}), but the behavior is analogous to a
left--to--right unary operator from an \emph{operational} viewpoint, played
here by the \texttt{complex} datatype that accepts one input argument on the
right \texttt{1.1}:

\begin{Verbatim}[fontsize=\scriptsize,numbers=left,frame=single]
var _a = (complex)1.1;
@factotum.alert(_a*2);
\end{Verbatim}
\noindent Typecasting also features chained construction:

\begin{Verbatim}[fontsize=\scriptsize,numbers=left,frame=single]
var _h = (quaternion)(complex)1.1;
@factotum.alert(_h*2);
\end{Verbatim}
Typecasting can be ‘implicitly’ used (i.e., without variable identifiers) and
sequentially chained to return a field or call a method:

\begin{Verbatim}[fontsize=\scriptsize,numbers=left,frame=single]
var _r = ( (complex)(1,2) ).real;
var _sin_i = sin( (complex)(1,2) );
\end{Verbatim}

\subsection{The \texttt{\#include} pre-directive.}\label{subsection_include}
This \texttt{\#include} pre-directive embeds the contents of input files into
the calling code, at the same entry line of the preprocessor, according to the
pattern: \texttt{\#include "<filepath>"}. It can appear inside bodies of
functions, of loops, of decisional statements, or of iterable data structures
such as (\texttt{Array} or \texttt{\JSON}s). The major benefit consists in not
dealing with large chunks of code: smaller pieces are easier to be read,
managed and possibly used again elsewhere:

\begin{Verbatim}[fontsize=\scriptsize,numbers=none,frame=single]
class someclass {
    constructor (){}

    #include "members.js"
    #include "methods.js"
}
\end{Verbatim}
Both files are assumed to be save in the same folder as of the above code.
Namely, the file \texttt{members.js} includes
\begin{Verbatim}[fontsize=\scriptsize,numbers=none,frame=single]
var _m1 = "str", _m2 = 1;
\end{Verbatim}
\noindent and the contents of the \texttt{methods.js} are

\begin{Verbatim}[fontsize=\scriptsize,numbers=none,frame=single]
my_method_1(){ return this._m1; }
my_method_2(){ return this._m2; }
\end{Verbatim}
In general, the \texttt{\#include} pre-directive reads the contents of the
files matching the following syntax patterns, also featuring wildcards:

\begin{Verbatim}[fontsize=\scriptsize,numbers=none,frame=single]
//single file per call
#include "folders-path/filename"

//multiple files from the whole input folder
#include "folders-path/filename/"

//same pattern, but with wildcards variation
#include "folders-path/filename/*"

//all files in the folder, starting with the letter "g"
#include "folders-path/filename/g*"

//all files in the folder, ending with suffix ".js"
#include "folders-path/filename/*.js"
\end{Verbatim}

\section{Debug}
This section includes some features that are available only in debug mode,
which requires a longer compile time than the standard run.
\subsection{Duplicate switch...case conditions}
If the \texttt{switch ... case} statement has been developed in a long span of
time as well as it might possibly include a very long number of \texttt{cases}
too, it may happen that a same case has been mentioned twice. The following
feature detects whether such duplicates have occurred:

\begin{Verbatim}[fontsize=\scriptsize,numbers=none,frame=single]
switch(a)
{
    case 1+1:
        doSomething();
    break;
    case 1+1:
    doSomethingElse();
    break;
}
\end{Verbatim}

\section{Conclusions}
The goal of this transpiler project is to introduce new features that may be
useful in advancing Javascript coding or, at the very least, to fork this
language in the direction of a strongly typified programming language devoted
to front-end Web development, as well as to introduce new syntax patterns or
improve existing ones to have a more versatile version. All of our suggestions
are meant to supplement, rather than replace, the existing Javascript
programming paradigms.

\end{document}